\documentclass{desyproc}
\def\eslt{\not\!\!{E_T}}
\def\eslt{E_T^{\rm miss}}

\def\to{\rightarrow}

\def\bi{\begin{itemize}}
 \def\ei{\end{itemize}}

\def\c1p{C1^\prime}
\def\ta{\tilde a}
\def\tG{\tilde G}

\def\ta{\tilde a}

\def\tg{\tilde g}

\def\tq{\tilde q}

\def\tz{\tilde\chi^0}
\def\alt{\stackrel{<}{\sim}}
\def\agt{\stackrel{>}{\sim}}
\def\be{\begin{equation}}  
\def\ee{\end{equation}}  
\def\bea{\begin{eqnarray}}  
\def\eea{\end{eqnarray}}

\begin{document}
\title{SUSY dark matter and the LHC\footnote{Invited talk given at
5th Patras workshop on axions, WIMPs and WISPs, Durham, England, July 13-17, 2009}}

\newcommand\prd[3]{{\it Phys.\ Rev.\ }{\bf D #1} (#2) #3}
\newcommand\prep[3]{{\it Phys.\ Rept.\ }{\bf #1} (#2) #3}
\newcommand\prl[3]{{\it Phys.\ Rev.\ Lett.\ }{\bf #1} (#2) #3}
\newcommand\plb[3]{{\it Phys.\ Lett.\ }{\bf B #1} (#2) #3}
\newcommand\jhep[3]{{\it J. High Energy Phys.\ }{\bf #1} (#2) #3}
\newcommand\app[3]{{\it Astropart.\ Phys.\ }{\bf #1} (#2) #3}
\newcommand\apj[3]{{\it Astrophys.\ J. }{\bf #1} (#2) #3}
\newcommand\ijmpd[3]{{\it Int.\ J.\ Mod.\ Phys.\ }{\bf D #1} (#2) #3}
\newcommand\npb[3]{{\it Nucl.\ Phys.\ }{\bf B #1} (#2) #3}
\newcommand\epjc[3]{{\it Eur.\ Phys.\ J. }{\bf C #1} (#2) #3}
\newcommand\ptp[3]{{\it Prog.\ Theor.\ Phys.\ }{\bf #1} (#2) #3}
\newcommand\zpc[3]{{\it Z.\ Physik }{\bf C #1} (#2) #3}
\newcommand\cpc[3]{{\it Comput.\ Phys.\ Commun.}{\bf #1} (#2) #3}
\newcommand\mpla[3]{{\it Mod.\ Phys.\ Lett.}{\bf A #1} (#2) #3}
\newcommand\arnps[3]{{\it Ann.\ Rev.\ Nucl.\ Part.\ Sci.}{\bf  #1} (#2) #3}
\newcommand\ppnp[3]{{\it Prog.\ Part.\ Nucl.\ Phys.}{\bf  #1} (#2) #3}
\newcommand{\hepph}[1]{hep-ph/#1}
\newcommand{\astroph}[1]{astro-ph/#1}

\author{{\slshape Howard Baer}\\[1ex]
Dep't of Physics and Astronomy, University of Oklahoma, Norman, OK 73019}

\contribID{lindner\_axel}

\desyproc{DESY-PROC-2009-05}
\acronym{Patras 2009} 


\maketitle

\begin{abstract}
Weak scale supersymmetry is a highly motivated extension of the 
Standard Model that has a strong degree of support from data. It provides
several viable dark matter candidates: the lightest neutralino (a WIMP), 
the gravitino, and the axion/axino supermultiplet. The LHC turn-on is imminent.
The discovery of supersymmetry at the LHC will go a long way towards
establishing the nature of dark matter. I present arguments why 
mainly axion cold dark matter is a better fit for supersymmetric models than
neutralinos. I also argue that Yukawa-unified 
SUSY GUT theories based on $SO(10)$ with mixed axion/axino cold dark matter
are extremely compelling, and present distinctive signatures for 
gluino pair production at the LHC.
\end{abstract}

\section{Introduction}

Astrophysical evidence for the existence of dark matter is now overwhelming,
and comes from disparate sources: galactic clustering, 
galactic rotation curves, anisotropies in the CMB, microlensing,
large scale structure, to name a few. 
The dark matter clusters on large scales, and helps seed structure
formation in the universe. While the identity of the dark matter particle,
or particles, is unknown, we do know several of its properties:
it must be massive, non-relativistic (cold or warm), electric and
color neutral, and stable at least on cosmic time-scales.

Of all the fundamental particles in the Standard Model (SM) of particle physics,
only the neutrinos come close to having these properties. However, neutrinos
are exceedingly light and engage in weak interactions: 
relic neutrinos would move at highly relativistic velocities, 
and so couldn't clump enough to seed structure formation. Their measured abundance 
from WMAP analyses is only a tiny fraction of the universe's energy budget.
Thus, the existence of dark matter is also evidence for physics
Beyond the Standard Model (BSM).

While there exist a plethora of candidate DM particles from BSM theories (examples include
black hole remnants, Q-balls, sterile neutrinos, axions, KK gravitons, gravitinos,
neutralinos, KK photons, branons and the lightest $T$-parity odd
particle of Little Higgs theories. Two of these stand out in that they arise naturally
due to very elegant solutions to long-standing problems in particle physics.
These include the {\it axion}, which arises from the Peccei-Quinn (PQ) solution to the strong 
CP problem of QCD\cite{pqww}, 
and the {\it lightest supersymmetric particle}, or LSP, of
$R$-parity conserving supersymmetric (SUSY) theories\cite{wss}. The SUSY theories solve 
the problematic quadratic divergences associated with scalar fields by introducing a 
new symmetry which relates bosons to fermions, thus giving scalar fields the milder
divergence structure which is held by chiral fermions and gauge fields. 
SUSY also provides a means to unification with gravity, is an essential part of 
superstring theory, and in concert with Grand Unified Theories (GUTs), 
receives some experimental support in the unification of gauge couplings under
MSSM renormalization group evolution.

In fact, the PQ strong CP solution and supersymmetry are in many ways made for each other,
so these two schemes are not mutually exclusive, and both may well be right. In that case, 
the axion occurs as part of an axion supermultiplet, which contains along with the axion an
$R$-parity odd {\it axino} $\ta$, which may serve as LSP. In SUSY theories, the
neutralino, the gravitino and the axino are all possible LSP candidates.
In this talk, I will restrict my comments to supersymmetric dark matter
(which includes axions and axinos), and comment on how it relates to LHC physics.  

\section{SUSY WIMP (neutralino) cold dark matter}

In SUSY theories with neutralino CDM, the $\tz_1$ is considered a natural
WIMP candidate for dark matter. The neutralino relic density can be calculated
by solving the Boltzmann equation as formulated for a FRW universe.
Central to the calculation is computation of the thermally averaged
neutralino annihilation (and co-annihilation) cross section.
The fact that the relic density comes out approximately in the right ball-park 
is often referred to as the ``WIMP miracle''.

The paradigm model for SUSY phenomenology is called minimal supergravity (mSUGRA)
or CMSSM. It is defined by just a few parameters: $m_0,m_{1/2},A_0,\tan\beta $ and $sign(\mu )$.
In the mSUGRA model, the WIMP miracle is actually no miracle at all! 
The relic density turns out to be much too large over most parameter space.
In fact, only in regions where the neutralino annihilation rate is highly enhanced 
will the relic density match the measured value. These regions are termed:
the stau co-annihilation region at low $m_0$, the HB/FP region at large $m_0$
where $\mu$ becomes small and we get mixed bino-higgsino CDM, the $A$-resonance
region at large $\tan\beta$ where $\tz_1\tz_1$ annihilation through the $A$-
resonance is enhanced, and the (largely excluded) bulk region, where
annihilation is through $t$-channel exchange of light sleptons.

Direct production of WIMP dark matter at LHC ({\it e.g.} $pp\to\tz_1\tz_1 X$) is usually not 
interesting since there is no hard energy deposition for detectors to trigger upon. 
However, if SUSY exists, then LHC may be able to produce many or all of the
{\it other} SUSY particles. The SUSY particle's subsequent cascade decays\cite{cascade} should lead
to collider events with high $p_T$ jets, high $p_T$ isolated $e$s and $\mu$s. Since
each sparticle cascade decay terminates at the LSP (the putative DM particle), the 
SUSY events should also contain large missing transverse energy ($\eslt$) due to 
non-detection of the DM particles. 

\begin{figure}[htb]
\centerline{\includegraphics[width=0.5\textwidth]{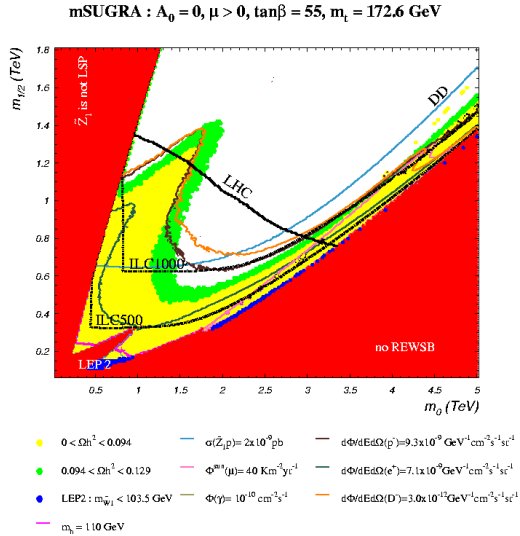}}
\caption{Contours of direct and indirect detection rates 
along with reach of LHC and iLC for SUSY in the mSUGRA model
for $\tan\beta =55$.
}\label{fig:sug}
\end{figure}

At each point in mSUGRA (or any other SUSY model) parameter space, we
can simulate LHC production of the entire array of superparticles, along with their
cascade decays\cite{lhc}. By looking for signals with high $p_T$ jets, isolated leptons and
$\eslt$-- beyond levels expected in the SM-- we can test if a signal can be seen for
an assumed value of integrated luminosity. In Fig. \ref{fig:sug}\cite{bbbkt,njp}, such a 
calculation has been made assuming 100 fb$^{-1}$ of integrated luminosity.
LHC should be able to see the parameter space below the contour marked ``LHC'',
which corresponds to $m_{\tg}\sim 3$ TeV when $m_{\tg}\sim m_{\tq}$, or
$m_{\tg}\sim 1.8$ TeV when $m_{\tq}\gg m_{\tg}$. We also show contours of
direct WIMP detection rates and indirect WIMP detection rates via high energy
neutrino detection at IceCube, or via detection of $\gamma$s, $e^+$s or $\bar{p}$s
arising from neutralino annihilation in the galactic halo.

It is noteworthy than in the DM-favored HB/FP region at large $m_0$, 
the LHC can only cover a portion of allowed parameter space. However, in this
region, direct detection via Xenon-100 or indirect detection via IceCube
is likely. In the $A$-annihilation region-- the large bump in the center of the plot-- 
detection of halo annihilations via $\gamma$, $e^+$ and $\bar{p}$ is enhanced\cite{bo}.

\begin{figure}[htb]
\centerline{\includegraphics[width=0.5\textwidth]{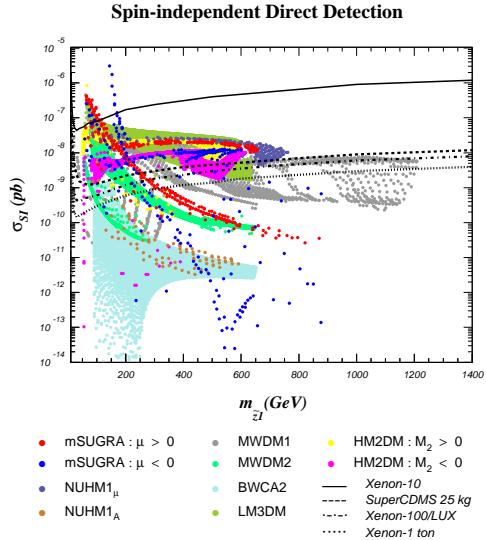}}
\caption{Direct detection rates for SUSY models with a well-tempered
neutralino. Each point represents a relic-density consistent model
with $\Omega_{\tz_1}h^2\simeq 0.11$.
}\label{fig:dd}
\end{figure}

The enhancement of DD and IDD in the higgsino-like region is a general feature of
a large assortment of models going beyond mSUGRA. In Fig. \ref{fig:dd}, we show
predicted rates in models with a {\it well-tempered} neutralino\cite{ag}. 
The large cluster of models
around $10^{-8}$ pb shows that the next set of DD experiments can either discover or
rule out an entire class of well-motivated SUSY models\cite{wtn}.

\section{The gravitino problem for WIMP and gravitino dark matter}

A potential pit-fall in the mSUGRA model is known as the gravitino 
problem. If we assume a SUGRA-type model, with a TeV scale
gravitino $\tG$, 
then gravitinos can be produced thermally in the early universe
(even though they are never in thermal equilibrium). If the $\tG$
is not the LSP, then it will decay into particle-sparticle pairs, 
and the sparticle cascade decays will contribute additional LSPs
to the relic density. The relic density is too much if the re-heat temperature
$T_R\agt 10^{10}$ GeV. Even if $T_R$ is lower, the late-time gravitino 
decays inject high energy particles into the cosmic soup during or after BBN,
which can destroy the successful BBN predictions which match so well with data.
Detailed calculations\cite{moroi} show that one needs $T_R\alt 10^5$ GeV
(which conflicts with many baryogenesis mechanisms) or $10^5<T_R<10^9$ GeV
as long as $m_{\tG}\agt 5$ TeV (the large $m_{\tG}$ suppresses the
gravitino lifetime to less than 1 sec, so $\tG\to f\tilde{f}$ decays
occur at the onset or even before BBN starts). Since in SUGRA models
the gravitino mass sets the scale for all the SSB terms, then we
would naively expect all the SUSY particles to be at masses $>5$ TeV
(beyond LHC reach).

One way out might be to assume the gravitino is the LSP.
But this results in the gravitino problem in reverse. Then
neutralinos or other sparticles in the early universe would decay
with long lifetimes in particle-gravitino pairs, and again disrupt
BBN. Detailed calculations\cite{moroi} show that $m_{\tG}$ should be
$\alt 1$ GeV for $m_{\tz_1}\sim 100-1000$ GeV. Since $m_{\tG}$
sets the scale for the other sparticle, we would expect all the
sparticles to have mass $\alt 1$ GeV, in contradiction to experimental limits.

\section{Mainly axion CDM in minimal supergravity model}

Another possibility is to assume some form of PQ solution to the strong CP
problem. In the SUGRA context, we will add an axion supermultiplet
to the model, which also contains an $R$-odd spin $1\over 2$ axino $\ta$\cite{rtw,roszk}.
Then, axions will be produced as usual via vacuum mis-alignment\cite{as},
and can contribute to the relic density. Their relic abundance depends on
the PQ breaking scale $f_a$, or alternatively on $m_a$.
A value of $f_a\sim 10^{12}$ GeV, corresponding to $m_a\sim 10^{-6}$ eV, 
would saturate the measured DM abundance. 

The value of the axino mass $m_{\ta}$ is very model-dependent: 
estimates range from the MeV to the multi-GeV scale\cite{roszk}. In mSUGRA, if
$\ta$ is the LSP, then $\tz_1\to \ta\gamma$ can occur with a lifetime of order
a fraction of a second: it is then BBN-safe. The $\ta$ can also be produced
thermally like the gravitino\cite{bsteffen}. Its thermal abundance depends on $f_a$,
$m_{\ta}$ and $T_R$. Thus, DM would have three components:
cold axions, cold axinos from thermal production and warm axinos
from neutralino decay.

\begin{figure}[htb]
\centerline{\includegraphics[width=0.5\textwidth]{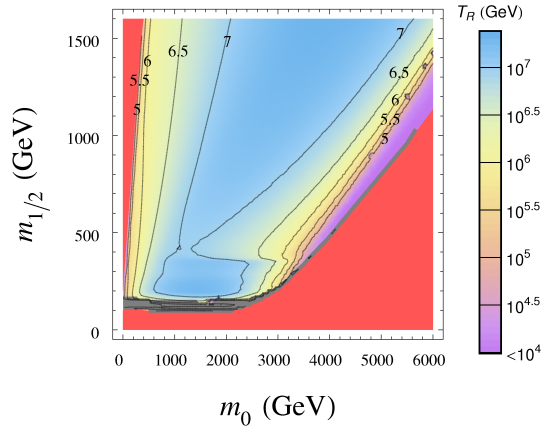}}
\caption{Contours of $\log_{10}T_R$ (the re-heat temperature)
in the mSUGRA model with mainly axion CDM, for $\tan\beta =10$.
}\label{fig:TR}
\end{figure}

The scenario works well if $m_{\ta}\sim $ the MeV scale.
Then, we find the mSUGRA model can have dominant axion CDM
with a small mixture of warm and cold axinos\cite{bbs}. The resulting value of $T_R$
is plotted in mSUGRA space in Fig. \ref{fig:TR}. The regions of mSUGRA
space that are most neutralino dis-favored lead to the highest values
of $T_R>10^6$ GeV. This is enough to sustain non-thermal or
Affleck-Dine leptogenesis! Thus, the most dis-favored neutralino
DM regions are precisely the {\it most favored} axion/axino
regions! Also, consequently, we expect quite different collider
signatures at LHC in the case of mixed axion/axino CDM, as compared
to neutralino DM.
We remark here that calculating the amount of fine-tuning of the neutralino 
relic abundance also shows a preference for mixed axion/axino CDM over neutralino CDM\cite{bb}.

\section{Yukawa-unified SUSY, mixed axion/axino CDM and the LHC}

SUSY GUT models based on the gauge group $SO(10)$ are extremely compelling, 
since these models allow for matter unification into the 16 dimensional spinor of
$SO(10)$, give rise naturally to see-saw neutrinos, and yield automatic cancellation
of triangle anomalies\cite{so10}.
The simplest $SO(10)$ SUSY GUT models also contain 
$t-b-\tau$ {\it Yukawa coupling unification}. Scans over SUSY model parameter 
space reveal that Yukawa unification only occurs for a very specific
spectra: first/second generation scalars at the 10 TeV level, third gen. scalars
and Higgs at the TeV scale, while gauginos are quite light, with 
$m_{\tg}\sim 300-500$ GeV and $m_{\tz_1}\sim 50-80$ GeV\cite{abbbft,bkss}. The neutralino relic
abundance turns out to be much too high: $\Omega_{\tz_1}h^2\sim 10^3 -10^4$, 
about 4-5 orders of magnitude too much. The problem can be solved by invoking
mixed axion/axino CDM\cite{bkss}. Then, $\tz_1\to \ta\gamma$ decays reduce the relic abundance
by factors of $10^4-10^5$. The scenario works best if the CDM is mainly axions, with
a small admixture of thermal and non-thermal axinos\cite{bhkss}. 
Values of $T_R\sim 10^7-10^8$ are
possible, allowing for baryogenesis. And since $m_{sparticle}\sim 10$ TeV, we also expect
$m_{\tG}\sim 10$ TeV, thus solving the gravitino problem!

In this model, the light gluinos lead to robust signatures at both the Tevatron\cite{tev}
and LHC colliders\cite{so10lhc}, and the whole scenario should be largely tested within 
year 1 of LHC operation, or even sooner if a Tevatron analysis is performed.






\end{document}